\documentclass[10pt,aps,pra,twocolumn]{revtex4-1}
\usepackage{bm}
\usepackage{amsmath}
\usepackage{amssymb}
\usepackage{graphicx}
\usepackage[latin1]{inputenc}
\usepackage[T1]{fontenc}
\usepackage{latexsym}
\usepackage{times}

\newcommand{\ket}[1]{|#1\rangle}

\newcommand{\uu}{\boldsymbol{u}}

\begin{document}

\title{Directional nanophotonic atom--waveguide interface based on spin--orbit interaction of light}

\author{R. Mitsch}
\author{C. Sayrin}
\author{B. Albrecht}
\author{P. Schneeweiss}
\author{A. Rauschenbeutel}
\affiliation{%
 \mbox{Vienna Center for Quantum Science and Technology, Atominstitut, TU Wien, Stadionallee 2, 1020 Vienna, Austria}
}

\maketitle

\bfseries Optical waveguides in the form of glass fibers are the backbone of global telecommunication networks~\cite{Kao10}. In such optical fibers, the light is guided over long distances by continuous total internal reflection which occurs at the interface between the fiber core with a higher refractive index and the lower index cladding. Although this mechanism ensures that no light escapes from the waveguide, it gives rise to an evanescent field in the cladding. While this field is protected from interacting with the environment in standard optical fibers, it is routinely employed in air- or vacuum-clad fibers in order to efficiently couple light fields to optical components or emitters using, e.g., tapered optical fiber couplers~\cite{Armani03,Morrissey13}. Remarkably, the strong confinement imposed by the latter can lead to significant coupling of the light's spin and orbital angular momentum~\cite{Zhao07,Hosten08,Bliokh12a}. Taking advantage of this effect, we demonstrate the controlled directional spontaneous emission of light by quantum emitters into a sub-wavelength-diameter waveguide. The effect is investigated in a paradigmatic setting, comprising cesium atoms which are located in the vicinity of a vacuum-clad silica nanofiber~\cite{Vetsch10}. We experimentally observe an asymmetry higher than 10:1 in the emission rates into the counterpropagating fundamental guided modes of the nanofiber. Moreover, we demonstrate that this asymmetry can be tailored by state preparation and suitable excitation of the quantum emitters. We expect our results to have important implications for research in nanophotonics and quantum optics and for implementations of integrated optical signal processing in the classical as well as in the quantum regime.

\mdseries 
It is well known that the radiation pattern of an ensemble of emitters can be modified and the directivity of the emitted light can be increased, e.g.,~by periodically arranging the emitters such that they fulfill a Bragg condition. Since the discovery of superradiance, it also became clear that ensembles without a long-range order can show a directed emission when they are suitably excited~\cite{Rehler71,Scully06}. Another way to influence the spontaneous emission characteristics of emitters consists in taking advantage of the Purcell effect in an optical resonator~\cite{Purcell46}. All the above experiments can be performed in the paraxial regime. In this case, the electromagnetic fields are transversally polarized and spin and orbital angular momenta of light are independent physical quantities. For example a circularly polarized collimated Gaussian laser beam has a well-defined spin of precisely one quantum of angular momentum $\hbar$ per photon but zero orbital angular momentum, while a linearly polarized collimated Gauss-Laguerre beam has zero spin but carries an integer multiple of $\hbar$ of orbital angular momentum per photon~\cite{Allen92}. In recent years, nanophotonic devices have gained increasing importance for many applications~\cite{Noda07,Novotny12}. In these structures, the light is strongly confined at the wavelength or sub-wavelength scale and, consequently, generally exhibits a significant spin--orbit interaction (see \cite{Bliokh12a} and references therein). 
Several experiments which probe the extraordinary angular momentum properties of spin--orbit coupled light have been suggested~\cite{Bliokh14}. As an example, metallic optical nanostructures provide such a strong field confinement and the directional launching of surface plasmon polaritons has been demonstrated~\cite{Rodriguez-Fortuno13,Lin13}. More recently, asymmetric scattering patterns of radio frequency waves based on spin--orbit interaction have been observed in sub-wavelength hyperbolic metamaterials~\cite{Kapitanova14}.

\begin{figure*}
	\includegraphics[width=1.0\textwidth]{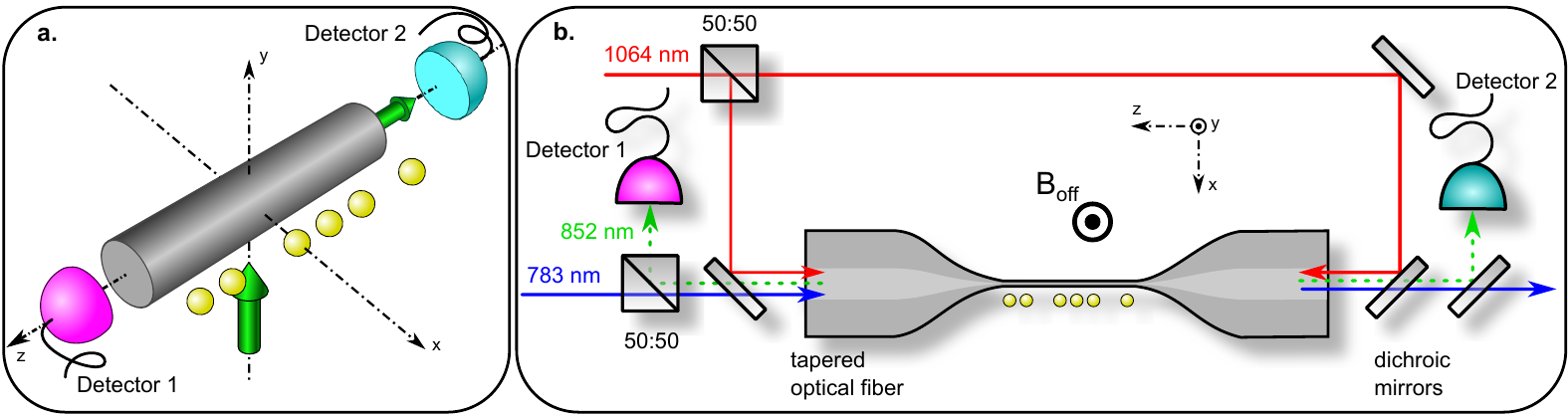}
	\caption{\textbf{a}: Schematic of the experiment: Atoms (yellow spheres) are trapped on one side of the nanofiber (radius $a$) at the transverse position ($y=0$ and here $a<x\approx 480~{\rm nm}$). A $\sigma^-$-polarized external laser beam (vertical green arrow) that propagates in the $+y$ direction excites the atoms. The fluorescence light that the atoms emit into the nanofiber is recorded using two detectors, one at each end of the fiber. \textbf{b}: Sketch of the setup including the tapered optical fiber (TOF), the dipole trap laser beam paths (red and blue lines), the resonant light beam paths (green dotted lines), and the atoms at the nanofiber section of the TOF. The orientation of the external homogeneous offset magnetic field is indicated. 50:50: balanced polarization-independent beam splitter. The wavelengths of the light fields are indicated.}
\label{fig:setup} 
\end{figure*}

Here, we demonstrate that spin--orbit interaction of light leads to directional spontaneous emission of photons by atoms into a nanophotonic waveguide. We use a small number of cesium atoms as quantum emitters. The atoms are located in the vicinity of the surface of a subwavelength-diameter silica nanofiber. Thanks to this close proximity, the atoms are efficiently interfaced with the waveguide modes via their evanescent field part. Consequently, a fraction of the atomic fluorescence couples into the waveguide. A sketch of the experiment is shown in Fig.~\ref{fig:setup}~\textbf{a}. As a key result of the present work, we find that more than 90~\% of the optical power that is emitted into the fundamental mode of the nanofiber can be launched into a given direction. The asymmetry depends on both, the position of the atoms relative to the waveguide and on the polarization of the light emitted by the atoms, which we control by properly choosing the atom's internal state and the polarization of the excitation laser light. The experiment is implemented using a nanofiber-based optical dipole trap for laser-cooled atoms~\cite{Vetsch10}. The trapping potential consists of two diametric linear arrays of individual trapping sites along the nanofiber, located $230$~nm from the surface. Each site contains at most one atom and provides a strong sub-wavelength confinement~\cite{Vetsch10} in every direction, considering the wavelength $\lambda=852$~nm of the atomic transition used in the experiment. In contrast to previous experiments performed with this system, only one linear array of atoms is prepared~\cite{ArXiv_Mitsch14}, see Methods. This allows us to locally probe the nanofiber-guided modes and to selectively place the atoms into regions of qualitatively different coupling. The optical nanofiber has a nominal radius of $a=250$~nm and is realized as the waist of a tapered optical fiber (TOF)~\cite{Brambilla10}. It enables almost lossless coupling of light fields that are guided in a standard optical fiber into and out of the nanofiber section. The experimental setup, including the TOF, the laser beam paths, and the trapped atoms, is shown in Fig.~\ref{fig:setup}~\textbf{b}. 

The physical origin of the directional spontaneous emission of light into the nanofiber lies in the polarization properties of the guided modes. For an atom at the position $\mathbf{r}$, the scattering rate into one of the nanofiber modes is proportional to ${\left|\mathbf{d}^*\cdot\uu(\mathbf{r})\right|}^2$, where $^*$ denotes the complex conjugation and $\mathbf{d}$ is the atomic dipole operator. The coupling between the atomic emitters and a nanofiber mode thus crucially depends on the local unit polarization vector $\uu$ of the latter~\cite{ArXiv_LeKien14}. For a sufficiently small fiber radius, as realized here, the optical nanofiber only guides the fundamental HE$_{11}$ mode~\cite{ArXiv_LeKien14}. These strongly guided optical fields are special in the sense that they show a significant coupling of the light's spin and orbital angular momentum~\cite{LeKien06b}. The electric part of the local spin density is proportional to the ellipticity vector, which is given by the cross product $i[\uu^*\times\uu]$.
 In strong contrast to paraxial light fields, the local spin density is position-dependent, in general not parallel to the guided field's propagation direction, and even orthogonal to it in the case of quasi-linearly polarized guided fields~\cite{Bliokh14,Bliokh12b,Reitz14}. Most importantly for the following, in the latter case, the local spin changes sign when reversing the propagation direction of the guided field. This effect is a clear signature of the coupling of the light's spin and orbital angular momentum. It allows us to control the direction of spontaneous emission that is coupled into the nanofiber.

\begin{figure*}
	\includegraphics[width=1.0\textwidth]{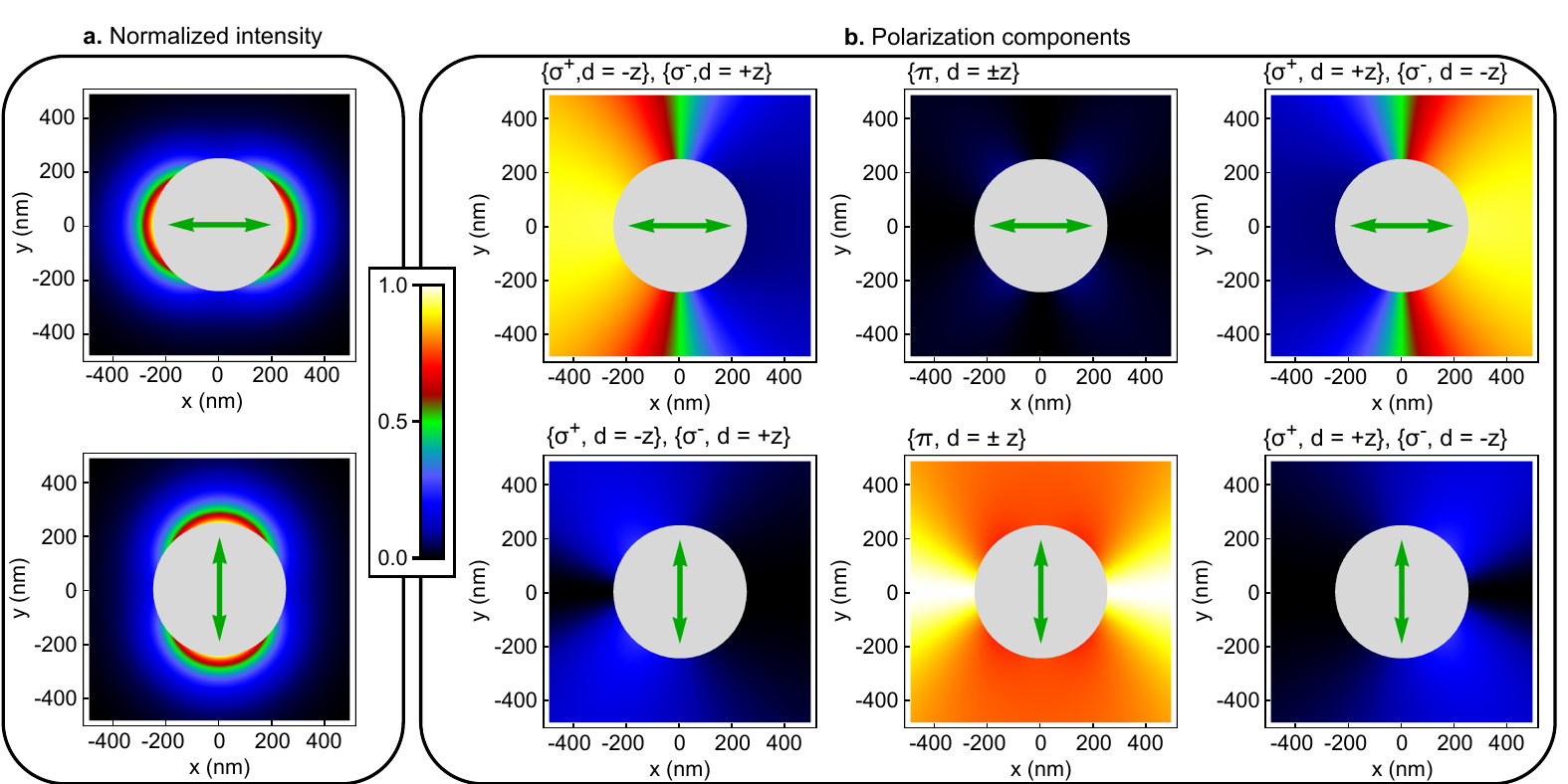}
	\caption{Nanofiber-guided quasi-linearly polarized basis modes. All quantities are calculated for running-wave fields with a wavelength of 852~nm and a 250~nm-radius silica fiber and are plotted in the plane transverse to the nanofiber axis. The orientation of the main polarization $p$ is indicated by a green double-arrow. \textbf{a}: Intensity distributions normalized to the maximal intensity at the nanofiber surface. A significant fraction of the guided optical power propagates outside the nanofiber, allowing one to efficiently interface atoms with the nanofiber-guided light field. The asymmetry of the intensity distribution in the azimuthal direction is apparent. \textbf{b}: Polarization overlaps, $\xi_{\sigma^+}$, $\xi_{\pi}$, and $\xi_{\sigma^-}$, for the specified propagation direction $d=\pm z$ of the nanofiber guided modes. The quantization axis is chosen along $+y$.}
	\label{fig:modes} 
\end{figure*}

In the following, we consider the quasi-linearly polarized HE$_{11}$ modes~\cite{ArXiv_LeKien14}. Four such modes, which have their main polarization $p$ oriented along the $x$-axis or along the $y$-axis ($p=x$ or $y$) and which propagate in the forward or backward propagation direction ($d=+z$ or $-z$), respectively, form a basis. The intensity of the quasi-linearly polarized basis modes is shown in Fig.~\ref{fig:modes}~\textbf{a}. Figure~\ref{fig:modes}~\textbf{b} shows a decomposition of the nanofiber-guided basis modes into the $\sigma^+$, $\pi$, and $\sigma^-$ polarization components~\cite{ArXiv_LeKien14}. Here, we take the $y$-axis as the quantization axis. With this choice, a $\sigma^+$- or $\sigma^+$-polarized light field exhibits a transverse spin. We plot the overlaps $\xi_j = {\left|\boldsymbol{e}_j^*\cdot\uu\right|}^2$, $j\in(\sigma^+,\pi,\sigma^-)$, of the polarization vector $\uu$ with the  orthonormal basis vectors $\boldsymbol{e}_\pi = \boldsymbol{e}_y, \boldsymbol{e}_{\sigma^\pm} = \pm\left(\boldsymbol{e}_x\mp i \boldsymbol{e}_z\right)/\sqrt{2}$. These overlaps are constant along the nanofiber axis and vary only slowly in the radial direction. However, they strongly vary as a function of the azimuthal position around the nanofiber. The circular polarization components of the guided modes, and thus the local spin density, depend on both, the position in the nanofiber transverse plane and the propagation direction of the mode. For the $p=x$ modes, at $(x<-a,y=0)$, i.e.,~on the left side of the nanofiber in Fig.~\ref{fig:modes}, $\xi_{\sigma^+}$ ($\xi_{\sigma^-}$) is maximal, when the propagation direction of the mode is $-z$ ($+z$), see upper left panel of Fig.~\ref{fig:modes}~\textbf{b}. At a distance of $230$~nm from the nanofiber surface ($x=-480$~nm), $\xi_{\sigma^+}=92~\%$ ($\xi_{\sigma^-}=92~\%$). Thus, these quasi-linearly polarized modes are locally almost perfectly circularly polarized, corresponding to a significant local spin density. Remarkably, this local spin points along the $y$-axis, i.e., is orthogonal to the propagation direction of the mode, and changes sign when the propagation direction is reversed. At $(x=+480\text{ nm},y=0)$, i.e., on the right side of the nanofiber in Fig.~\ref{fig:modes}, $\xi_{\sigma^+}$ ($\xi_{\sigma^-}$) is only 8\%, when the propagation direction of the mode is $-z$ ($+z$): The local spin density has opposite signs on opposite sides of the nanofiber. This effect is often referred as spin-Hall effect of light~\cite{Onoda04,Hosten08,Bliokh08a}. The overlap $\xi_\pi$, however, does not show a dependence on the propagation direction and is identical on opposite sides of the nanofiber. For the $p=x$ modes, $\xi_{\pi} \leq 8\%$ and, in particular, $\xi_{\pi}=0$ along the line $(x,y=0)$. The $p=y$ modes contain no circular polarization ($\xi_{\sigma^\pm}=0$ and $\xi_\pi=1$) along $y=0$. Thus, along this line, the polarization of the $p=y$ modes, and so the local spin density, is independent of their propagation directions.

For our choice of quantization axis, our discussion revealed that the dependence of the local polarization on the propagation direction is strongest in the plane $y=0$. In order to experimentally characterize the spontaneous emission of atoms into the optical nanofiber, we thus position the atoms either at $(x \approx 480~{\rm nm},y=0)$ or at $(x\approx -480~{\rm nm},y=0)$. At these positions, the two $p=y$ modes are exactly $\pi$-polarized and the $p=x$ modes are almost circularly polarized. The sign of the circularity is opposite for opposite propagation directions or on opposite sides of the nanofiber. As a consequence, a $\pi$-polarized photon emitted by an atom couples exclusively and equally to the two counter-propagating $p=y$ modes, while a $\sigma^\pm$-polarized photon preferentially couples to one of the two $p=x$ modes.

\begin{figure*}
\centerline{\includegraphics{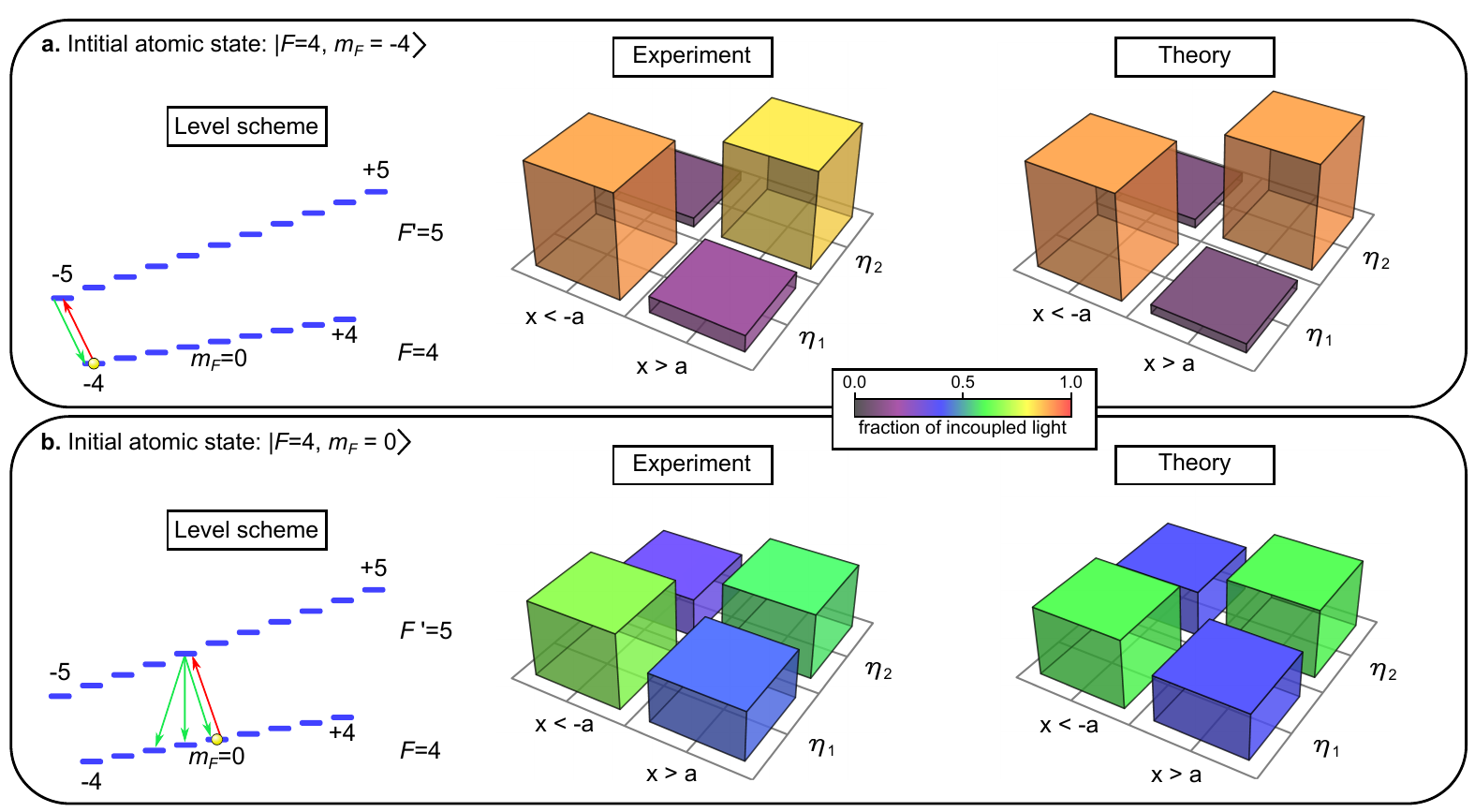}}
  \caption{Directed spontaneous emission into an optical nanofiber. Left: Atomic level scheme indicating the initial atomic state (yellow sphere), the transition driven by the external excitation laser field (red arrow) and the decay channels (green arrows). Center: Measurement results. Right: Theory prediction. \textbf{a}: The initial atomic state is $\ket{F=4, m_F=-4}$ state. The position of the atoms ($x\approx -480~{\rm nm}$ or $x\approx 480~{\rm nm}$) determines into which direction light is preferentially coupled into the nanofiber. More than 90~\% of the incoupled light propagates into a given direction. \textbf{b}: The initial atomic state is $\ket{F=4, m_F=0}$. The emission of the atom into the nanofiber is almost balanced. The experimental data in panels \textbf{a} and \textbf{b} are averaged over 3200 and 4000 experimental runs, respectively. The experimental data and theory are in good agreement.}
\label{fig:directional}
\end{figure*}%

For the first set of measurements, we prepare the atoms in the outermost Zeeman substate $\ket{F=4, m_F=-4}$. Here, $F$ is the total angular momentum quantum number and $m_F$ the magnetic quantum number. In order to avoid spin flips~\cite{ArXiv_Mitsch14} and to spectrally separate neighboring optical transitions, a magnetic offset field of $28$~G is applied in the $y$-direction, i.e., along the quantization axis. We drive the atomic cycling transition $\ket{F=4, m_F=-4} \to \ket{F'=5, m_{F'}=-5}$ using an external $\sigma^-$-polarized laser field on the D2 line. At the position of the atoms, the polarization of the  excitation laser is not modified by the nanofiber, see Supplemental Material. The involved atomic levels and transitions are shown in the left panel of Fig.~\ref{fig:directional}~\textbf{a}. On this closed transition, the atoms only emit $\sigma^-$-polarized light. Two detectors, one at each end of the tapered optical fiber, are used to record the number of photons that are coupled into the nanofiber, see Fig.~\ref{fig:setup}. The measurement interval is limited by the drop of the signal after about $10\ \mu$s which we attribute to the loss of atoms from the trap due to photon recoil heating. We sum up all recorded photon counts individually for each detector and correct for the optical losses of the setup. In the center panel of Fig.~\ref{fig:directional}~\textbf{a}, we plot the fractions $\eta_1$ and $\eta_2$ of the total incoupled nanofiber-guided light that was detected by detector 1 and 2, respectively. When the atomic sample is located at $x\approx -480~{\rm nm}$, detector 1 (receiving light that propagates in the $+z$ direction) records a significantly larger signal ($\eta_1=92\ \%\pm3\ \%$) than detector 2. The main propagation direction of the incoupled light is reversed when the atomic sample is prepared on the other side of the nanofiber, $x\approx 480~{\rm nm}$. In this case, detector 2 records the largest signal ($\eta_2=86\ \%\pm3\ \%$). For this measurement configuration, a straightforward quantitative estimation of the expected fraction $\eta_1=1-\eta_2$ can be made and is shown in the right panel of Fig.~\ref{fig:directional}~\textbf{a}: Only $\sigma^-$-polarized light is emitted and the coupling of the atoms to the locally $\pi$-polarized $p=y$ modes is thus zero. Hence, $\eta_1$ is simply given by the position-dependent polarization overlap $\xi_{\sigma^-}$ of the $p=x$ mode that propagates towards detector 1. This overlap is $92\%$ ($8\%$) when the atoms are located at $x=-480~\rm{nm}$ ($x=+480~\rm{nm}$), which is in very good agreement with our experimental results.

We now prepare the atomic sample in the $\ket{F=4, m_F=0}$ state. The external $\sigma^-$-polarized light field excites the atoms into the $\ket{F'=5, m_{F'}=-1}$ state, which can spontaneously decay via a $\sigma^-$, $\pi$ or $\sigma^+$ transition, leading to the emission of a photon with the corresponding polarization. The involved atomic levels and transitions and the experimental results are shown in Fig.~\ref{fig:directional}~\textbf{b}. Compared to the situation where we prepared the $\ket{F=4,m_F=-4}$ state, the emission into the nanofiber is now almost balanced. We find $\eta_1=66\ \%\pm2\ \%$ ($\eta_2=57\ \%\pm1\ \%$) for the atoms at $x\approx -480~{\rm nm}$ ($x\approx 480~{\rm nm}$). These smaller values are theoretically expected: The probabilities for the emission of $\sigma^-$, $\pi$ and $\sigma^+$ light for a decay from the $\ket{F'=5, m_{F'}=-1}$ state are $P_{\sigma^+}=2/15$, $P_{\pi}=8/15$, and $P_{\sigma^-}=5/15$, respectively. As already discussed, the emitted $\pi$-polarized light couples symmetrically into the waveguide. This light thus yields the same signal on the two detectors and reduces the contrast of any directed emission into the nanofiber. Moreover, as $\sigma^-$- and $\sigma^+$- polarized photons are emitted with similar probabilities, the emission rates into the counter-propagating modes of the nanofiber are almost equal. Our calculations then predict that $60\%$ of the total emission coupled into the nanofiber propagate into one direction~\cite{ArXiv_LeKien14}. Here, we also take into account the fact that the intensities of the $x$- and $y$-polarized nanofiber modes are not equal at the position of the atoms~\cite{Reitz14}, see Fig.~\ref{fig:modes}~\textbf{a}. Our prediction is in good agreement with the experimental results.

In conclusion, we employed spin--orbit interaction of light to realize a directional nanophotonic atom--waveguide interface. We carried out our experiments with cesium atoms in the vicinity of a silica optical nanofiber. We demonstrated that the emission into the nanofiber in a given direction can be more than ten times stronger than in the opposite direction. By preparing the atoms in different internal Zeeman substates, we showed that the coupling ratios can be controlled via the polarization of the emitted light. Our work thus highlights how spin--orbit interaction of the nanofiber-guided light fundamentally influences the spontaneous emission process.

The presented effects are universal in the sense that they should also occur for other strongly-confined optical fields~\cite{Noda07,Novotny12}, e.g., in integrated photonic waveguides~\cite{Lund-Hansen08}. In the view of the rise of technologies such as silicon photonics~\cite{Graydon10}, we therefore expect our findings to have an important impact on integrated optical signal processing. Our observations also pave the way towards an atom-mediated quantum photon router, in which the state of an atom controls the propagation direction of guided optical photons and which might thus constitute a central component for an optical quantum network~\cite{Kimble08}.

In the course of completion our manuscript, we became aware of two related theoretical works~\cite{ArXiv_Young14,ArXiv_Soellner14}. In both references a directional interface between a quantum dot and a photonic-crystal waveguide, that relies on spin--orbit interaction of light, is proposed.

\section*{Methods}
\subparagraph*{\hskip-10pt Details on the nanofiber-based two-color trap and the prepared atomic sample}
\footnotesize
Laser-cooled cesium atoms are trapped in the evanescent field surrounding a silica optical nanofiber of nominal radius $a = 250$~nm. The trapping potential is created by sending a blue-detuned running wave field with a free-space wavelength of $783$~nm and a power of $8.5$~mW and a red-detuned standing wave field at $1064$~nm wavelength with a power of $0.77$~mW per beam into the nanofiber~\cite{Vetsch10}. The blue- and the red-detuned fields are guided as quasi-linearly polarized fundamental HE$_{11}$ modes. The main polarizations of the two fields are perpendicular to each other. Two diametric arrays of trapping sites are formed, and the calculated radial, azimuthal, and axial trap frequencies of each site are $120$, $87$, $186$~kHz, respectively. The trap minima are located at a distance of $230$~nm away from the nanofiber surface.

The atoms are loaded into the nanofiber-based trap from a magneto-optical trap via an optical molasses stage~\cite{Vetsch10}. In this process, the collisional blockade effect limits the maximum number of atoms per trapping site to one and results in a maximum average filling factor of $0.5$~\cite{Vetsch10}. After loading, atoms are distributed over the two diametric arrays of trapping sites. For the study of the spontaneous emission into the nanofiber guided modes, the atoms in one of the two diametrically arranged arrays have to be removed. Otherwise, the symmetry of the system would prevent us from observing the directional emission into the nanofiber. In order to selectively remove atoms from one array of the nanofiber-based trap, we take advantage of a recently demonstrated technique for the preparation of atoms in one specific Zeeman state on one side of the nanofiber~\cite{ArXiv_Mitsch14}. We end up with a few tens of atoms in a given state $\ket{F=4, m_F}$ at either $x\approx -480$~nm or $x\approx 480$~nm, i.e., on the left or the right side of the fiber in Fig.~\ref{fig:modes}.

\bigskip

\section*{Acknowledgements}
\noindent Financial support by the Austrian Science Fund (FWF, SFB NextLite Project No.~F 4908-N23 and DK CoQuS project No.~W~1210-N16) and the European Commission (IP SIQS, No.~600645) is gratefully acknowledged. C.S.~acknowledges support by the European Commission (Marie Curie IEF Grant 328545).

\section*{Author Contributions}
\noindent R.M. and C.S. equally contributed to this work. R.M., C.S. and B.A. performed the experiment. R.M., C.S and P.S. analyzed the data. R.M., C.S., P.S. and A.R. wrote the manuscript. All authors discussed the results and reviewed the manuscript.

\section*{Additional information}
\noindent The authors declare no competing financial interests. Correspondence and requests for materials should be addressed to P.S.~or A.R.~(email: schneeweiss@ati.ac.at, arno.rauschenbeutel@ati.ac.at).

\bibliographystyle{nature}
\bibliography{Directional}

\end{document}


\title{Supplementary information to\\``Directional nanophotonic atom--waveguide interface based on spin-orbit interaction of light''}

\author{R. Mitsch, C. Sayrin, B. Albrecht, P. Schneeweiss, and A. Rauschenbeutel}
\affiliation{Vienna Center for Quantum Science and Technology, Atominstitut, TU Wien, Stadionallee 2, 1020 Vienna, Austria}

\maketitle

\section{Preparation of atoms}
\subsection{Preparing the atoms on one side of the fiber in the state $\ket{F=4, m_F=0}$}
Cesium atoms are cooled and spatially confined in a magneto-optical trap around the nanofiber. After loading them into the nanofiber-based two color dipole trap~\cite{LeKien04}, the atoms are arranged in two diametric linear arrays along the fiber~\cite{Vetsch12}. Initially they are in the $F=4$ manifold. After applying a magnetic offset field $B_{\text{off}}=3$~G along the $y$ axis to avoid spin flips, we optically pump the atoms in the $\ket{F=4, m_F=0}$ state. For the next steps we increase the offset field to 28~G. All the atoms are optically pumped to $F=3$: Most of the atoms are now in $\ket{F=3, m_F=0}$ whereas $\ket{F=4}$ is empty. To address the atoms on the different sides of the fiber separately, we send a light field at a tune-out wavelength (880~nm) into the fiber. It energetically shifts the $\ket{F=4, m_F=0} \rightarrow \ket{F=3, m_F=0}$ clocktransition differently for the atoms on different sides of the fiber~\cite{Arxiv_Mitsch14}. Here, the two transition frequencies were separated by $\approx 16$\,kHz. Finally, the atoms on one side of the fiber are transferred with a 14.7~$\mu$s microwave (MW) $\pi$-pulse from the $\ket{F=3, m_F=0}$ to the $\ket{F=4, m_F=0}$ state. Thus, most of the atoms of one specific side of the fiber are in $\ket{F=4, m_F=0}$ and all the other Zeeman sub-states in the $F=4$ manifold are not populated. The atoms trapped on the other side of the nanofiber were not addressed by the MW $\pi$-pulse and remain in the $F=3$ manifold. Then the experiments described in the main article can be performed by applying a $\sigma^-$-polarized laser propagating along the quantization axis. It addresses the atoms on only one side of the fiber on the $\ket{F=4, m_F=0} \rightarrow \ket{F'=5, m_F=-1}$ transition.\\

\subsection{Preparing the atoms on one side of the fiber in the state $\ket{F=4, m_F=-4}$}
For preparing the atoms in the outermost Zeeman sub-state all the steps described in the paragraph before are done. In addition, before applying the external $\sigma^-$ polarized laser, the atoms on one side of the fiber that are in $\ket{F=4, m_F=0}$ are optically pumped to $\ket{F=4, m_F=-4}$~\cite{Arxiv_Mitsch14}. The atoms trapped on the other side of the nanofiber are in the $F=3$ state and thus not addressed by the optical pumping light field. This light field is fiber guided and $\sigma^-$ polarized at the position of the atoms. Its frequency is scanned over the $\ket{F=4}\rightarrow\ket{F'=5}$ transition in 1~ms. The frequency scan is necessary to address all the Zeeman substates that are splitted by the magnetic offset field $B_{\text{off}}=28$~G. After this procedure the atoms on one side of the fiber are in $\ket{F=4, m_F=-4}$ and the experiments described in the main article can be performed.

\begin{figure*}[t b]
\includegraphics[width=1.0\textwidth]{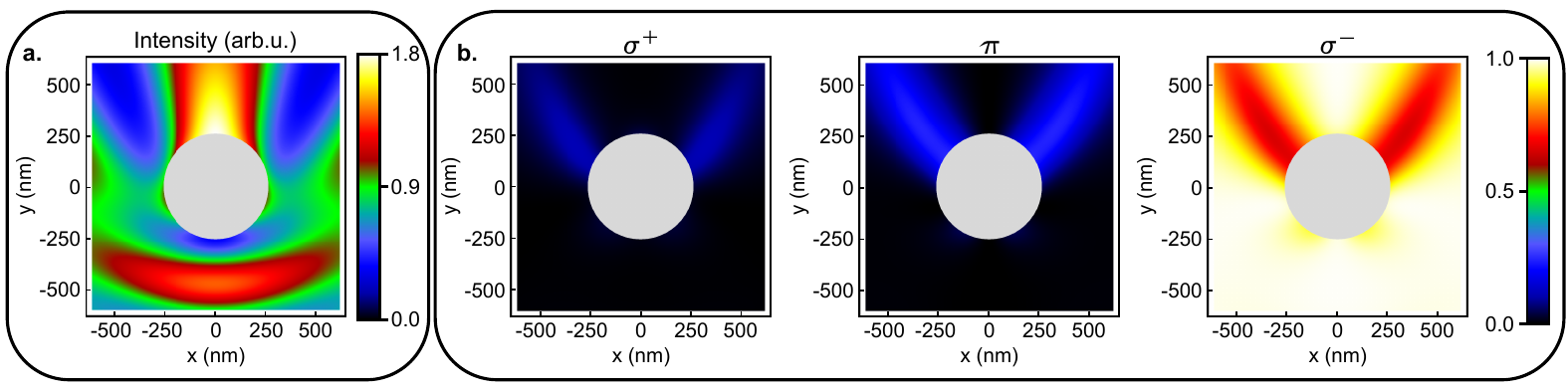}
  \caption{\textbf{a} Intensity and \textbf{b} polarization components  $\xi_{\sigma^-}$, $\xi_\pi$, and $\xi_{\sigma^+}$ around the nanofiber of a $\sigma^-$-polarized free-space plane wave propagating in the $+y$ direction. The intensity in \textbf{a} is normalized to the incident intensity. The quantization axis is chosen along $+y$, $a=250$~nm, $\lambda=852$~nm.}
\label{fig:side}
\end{figure*}

\section{Emission of an atom into nanofiber-guided optical modes}
\subsection{Quantitative estimations}
We estimate the expected fraction $\eta_1$ of the total incoupled nanofiber-guided light that propagates in the $+z$ direction for atoms prepared in a specific Zeeman substate $\ket{F=4,m_F}$ and excited by an external $\sigma^-$-polarized light field. The following calculations consider only one single atom. This is justified since the contributions of all atoms add up incoherently. An atom that has been excited by the external light field has in general three possible decay channels. The probability for the atom to decay via a $\sigma^+$, $\sigma^-$, or $\pi$ transition is $P_{\sigma^+}$, $P_{\sigma^-}$, or $P_{\pi}$, respectively. For the calculation of $\eta_1$ we calculate the overlap of the emitted polarization with the four basis modes. The basis modes have their main polarization $p$ oriented along the $x$-axis or along the $y$-axis ($p = x$ or $y$) and propagate in the forward or backward propagation direction ($d = +z$ or $-z$), respectively. The probability of scattering into the $+z$ direction can be written as
\begin{align}
	P(+z)&=P(+z,p=x)+P(+z,p=y)\nonumber\\
	&=P(+z,p=x|\sigma^+)\cdot P_{\sigma^+}+P(+z,p=x|\sigma^-)\cdot P_{\sigma^-}\nonumber\\
	&+\underbrace{P(+z,p=x|\pi)}_{=0}\cdot P_{\pi}\nonumber\\
	&+\underbrace{P(+z,p=y|\sigma^+)}_{=0}\cdot P_{\sigma^+}+\underbrace{P(+z,p=y|\sigma^-)}_{=0}\cdot P_{\sigma^-}\nonumber\\
	&+P(+z,p=y|\pi)\cdot P_{\pi}\,.
\end{align}
The probabilities $P(+z,p|\sigma^+)$, $P(+z,p|\sigma^-)$, and $P(+z,p|\pi)$that a $\sigma^+$, $\sigma^-$, or $\pi$ polarized photon, respectively, is emitted into the $p$ basis mode propagating in the $+z$ direction are given by the product of two conditional probabilities. One is given by the overlap of the polarization of the emitted photon with the $p$ basis mode, namely $\xi_{\sigma^+}$, $\xi_{\sigma^-}$, and $\xi_{\pi}$, respectively. The other contribution, denoted by $\alpha_x$ ($\alpha_y$), takes into account the intensity of the $p=x$ ($p=y$) basis mode at the position of the atom~\cite{Reitz14}. Thus, we get
\begin{align}
	P(+z)=0.92\cdot \alpha_x P_{\sigma^+}+0.08\cdot \alpha_x P_{\sigma^-}+\alpha_y\cdot P_{\pi}\,,
\end{align}
where the intensity ratio at the position of the atoms for the two modes is given by $\alpha_y/\alpha_x=0.36$.

For an atom that is prepared in the $\ket{F=4,m_{F}=0}$ state and then excited to the $\ket{F'=5,m_{F'}=-1}$ state we have~\cite{Steck10}: $P_{\sigma^-}=2/15$, $P_{\pi}=8/15$, and $P_{\sigma^+}=5/15$. Thus, in this case our calculation predicts that $\eta_1=P(+z)/(P(+z)+P(-z))=60\,$\% of the incoupled light is propagating in the $+z$ direction. For an atom prepared in the  $\ket{F=4,m_{F}=-4}$ state and then excited to the $\ket{F'=5,m_{F'}=-5}$ state the situation is even simpler, since $P_{\sigma^-}=1$, $P_{\pi}=0$, and $P_{\sigma^+}=0$ and thus $\eta_1=92\,$\%.

A generalized model is given in~\cite{ArXiv_LeKien14}. It describes the scattering of an atom into the nanofiber-guided fundamental HE$_{11}$ modes. With this model it is possible to calculate absolute scattering rates and to consider coherent superpositions of different $m_F$ states of an atom.

\section{Light scattering by an optical nanofiber at normal incidence}
In the experiment, the atoms trapped in the vicinity of the optical nanofiber are excited with an external light field. In this section, we study the effect of the optical nanofiber on the intensity and the local polarization of the excitation light field.
We follow the approach outlined in~\cite{Barber98} and approximate the incident laser light field by a plane wave that has its wave vector perpendicular to the nanofiber axis. For the calculations the nanofiber is described as a thin silica cylinder with a radius of $a=250$\,nm. We consider the incident light field to be $\sigma^-$ polarized with a wavelength of 852\,nm and propagating along the $+y$ axis, corresponding to the situation presented in the main text. The intensity and the overlaps $\xi_j = {\left|\boldsymbol{u}\cdot\boldsymbol{e}_j^*\right|}^2$, $j\in(\sigma^+,\pi,\sigma^-)$, of the polarization vector $\boldsymbol{u}$ with the basis vectors $\boldsymbol{e}_\pi = \boldsymbol{e}_y, \boldsymbol{e}_{\sigma^\pm} = \pm\left(\boldsymbol{e}_x\mp i \boldsymbol{e}_z\right)/\sqrt{2}$ are presented in Fig.~\ref{fig:side}.

As apparent from the figure, the alterations to the incident field can be, in general, significant. For example, the focusing of the incident field by the nanofiber leads to a two-fold increase of the field intensity at $x=0,\;y=250$~nm. Moreover, the polarization components  $\xi_{\sigma^-}$, $\xi_\pi$, and $\xi_{\sigma^+}$ vary around the fiber (see Fig.~\ref{fig:side} \textbf{b}). However, in our experiment, we only consider the two locations ($x=\pm (a+230)$~nm, $y=0$) where the two arrays of trapping sites are formed. Here, the deviations form the laser field free space intensity and polarization is small, such that it is justified to neglect the alterations of the nanofiber to the incident light field. In addition, it is clear from  Fig.~\ref{fig:side} that the intensity and all polarization components are modified in the same way at the two trapping site positions. 

\bibliography{Directional}
\bigskip